\documentclass[fleqn,10pt]{wlscirep}
\usepackage[utf8]{inputenc}
\usepackage[T1]{fontenc}
\usepackage{gensymb}
\usepackage{float}
\usepackage{subcaption} 
\usepackage{ulem}
\usepackage{pdfpages}

\title{Laser-induced thermal source for cold atoms}

\author[1]{Chung Chuan Hsu}
\author[2,3]{R\'emy Larue}
\author[1,3,*]{Chang Chi Kwong}
\author[1,3,4]{David Wilkowski}
\affil[1]{School of Physical and Mathematical Science, Nanyang Technological University, 637371 Singapore, Singapore}
\affil[2]{Laboratoire de Physique Subatomique et de Cosmologie, Universit\'e Grenoble-Alpes, CNRS/IN2P3, Grenoble INP, 38000 Grenoble, France}
\affil[3]{MajuLab, International Research Laboratory, IRL 3654, CNRS, Universit\'e C\^ote d'Azur, Sorbonne Universit\'e, National University of Singapore, Nanyang Technological University, Singapore.}
\affil[4]{Centre for Quantum Technologies, National University of Singapore, 117543 Singapore, Singapore}

\affil[*]{changchikwong@ntu.edu.sg}

\begin{abstract}
We demonstrate a simple and compact approach to laser cool and trap atoms based on laser-induced thermal ablation (LITA) of a pure solid granule. A rapid thermalisation of the granule leads to a fast recovery of the ultra-high vacuum condition required for a long trapping lifetime of the cold gas. We give a proof-of-concept of the technique, performing a magneto-optical trap on the 461~nm $^1S_0\rightarrow\,^1P_1$ transition of strontium. We get up to $3.5$ million of cold strontium-88 atoms with a trapping lifetime of more than $4$~s. The lifetime is limited by the pressure of the strontium-free residual background vapour. We also implement an original configuration of permanent magnets to create the quadruple magnetic field of the magneto-optical trap. The LITA technique can be generalized to other atomic elements such as transition metals and lanthanide atoms, and shows a strong potential for applications in quantum technologies ranging from quantum computing to precision measurements such as outdoor inertial sensing. 
\end{abstract}
\begin{document}

\flushbottom
\maketitle
\thispagestyle{empty}

\section*{Introduction}

Since most atomic metals used today for laser cooling and trapping have a low saturation pressure at room temperature, a typical atomic source consists of an effusive thermal atomic beam, which is extracted from a high temperature oven. Some examples of cold atoms produced from oven source include atomic species like alkaline-earth metals (Mg\cite{loo2003investigations}, Ca\cite{grunert2002sub}, Sr\cite{katori1999magneto,barbiero2020}, Ba\cite{de2009magneto}, Ra\cite{guest2007laser}), lanthanides (Eu\cite{inoue2018magneto,miyazawa21}, Dy\cite{youn2010dysprosium}, Ho\cite{miao2014magneto}, Er\cite{ilzhofer2018two}, Tm\cite{golovizin2019inner}, Yb\cite{maruyama2003investigation}), and transition-metals (Cr\cite{bradley2000magneto,beaufils2008all}). Due to the high longitudinal velocity of the hot atomic beam, a Zeeman slower is often required between the oven and the final magneto-optical trap (MOT). In some cases, a two-dimensional MOT is further inserted after the Zeeman slower to collimate and deflect the atomic beam, leading to an improvement in the loading rate of MOT and the vacuum environment in the science chamber \cite{yang2015high}. While the performance of such a system in terms of atoms number and lifetime is excellent, the experimental setups are bulky and usually require substantial maintenance efforts, which increases the challenges to operate cold atoms platforms outside dedicated laboratories \cite{poli2014transportable,yasuda2017laser, grotti2018geodesy,takamoto2020test}. 

Besides using an oven source, it is also possible to use laser ablation of a target sample as a source for neutral atoms~\cite{chu1985three,chu1986experimental,kim1997buffer,hemmerling2014}, ions~\cite{leibrandt2007,zimmermann2012,olmschenk2017,vrijsen2019,osada2021deterministic}, and molecules~\cite{tarallo2016,baum2020,mitra2020}. Recently, with the aim of simplifying the setups, laser-controlled sources have been developed for cold strontium~\cite{kock2016laser} and ytterbium~\cite{yasuda2017laser}. In these studies, the targets are oxides of strontium or ytterbium, and the dominant mechanism to release the atoms involves a photochemical process. This process has the advantage of requiring moderate laser power (\textit{i.e.} in the milliwatt range) but leads to an undesirable release of oxygen as a by-product, which contributes to a higher background pressure that limits the lifetime of the cold gas. 

In this report, we demonstrate an alternative approach that uses pure metallic granule as the ablation target. The pulse duration in our work  ranges from few tens of milliseconds to few seconds, where the duration is long enough for thermal process to occur~\cite{chichkov1996, zimmermann2012}. To indicate the nature of our ablation process, we will refer to it as laser-induced thermal ablation (LITA). In the LITA process, a focused ablation laser beam strongly heats a micrometre-sized region of the granule to produce an atomic vapour.  The MOT is then loaded from the vapour within the same vacuum chamber. The advantage of such an approach lies in the rapid reduction of the background vapour pressure once the ablation laser is turned off. Moreover, in contrast to the oxides-based techniques~\cite{kock2016laser,yasuda2017laser}, the LITA does not rely on specific chemical bonds, and can be implemented in principle to any pure solid state elements. We illustrate the operation of this simple method using strontium-88 atoms. With our ablation-MOT setup, we get up to $3.5(2)\times$10$^6$ cold atoms with a lifetime longer than $4$~s in a MOT operating on the $^1S_0\rightarrow\,^1P_1$ dipole-allowed strontium transition at $461$~nm. 


\section*{Methods}

\subsection*{Experimental setup}

\begin{figure}[H]
\centering
\includegraphics[width=0.8\linewidth]{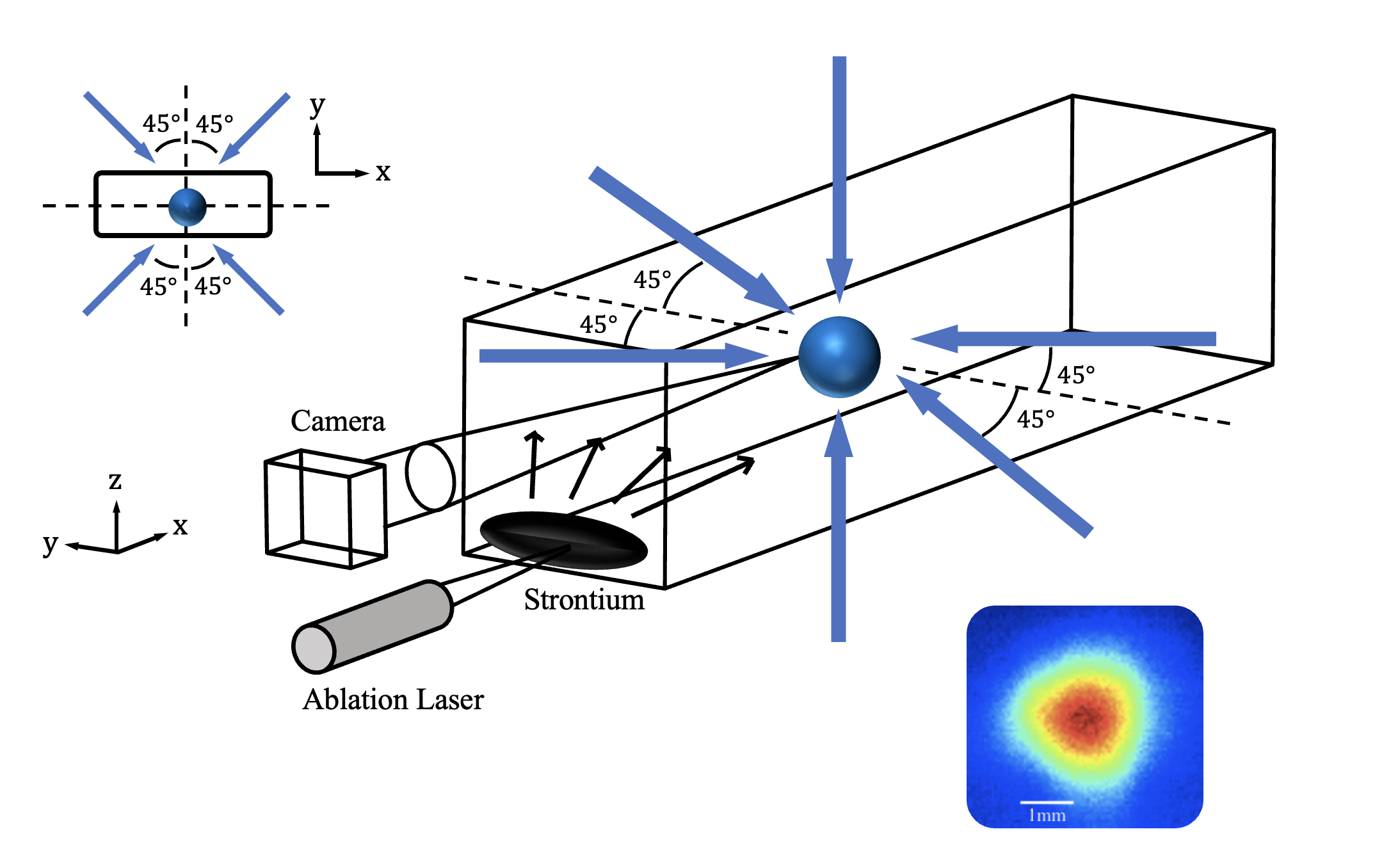}
\caption{A strontium MOT in a vacuum chamber (rectangular cuboid). The blue arrows represent the $461$~nm MOT beams. The magnetic  fields and repumper lasers are not shown here. \textit{Inset:} sample image of the cold atoms cloud rendered in false colour (size $\sim$~$1.5$~mm).}
\label{fig:exp setup}
\end{figure}

Figure \ref{fig:exp setup} shows a schematic of our experimental setup. The vacuum chamber is an uncoated borosilicate glass cell with dimensions 30~mm $\times$ 30~mm $\times$ 100~mm. A strontium granule is loaded into one end of the glass cell. After a bake-out of the vacuum system, we achieve an ultra-high vacuum condition that we sustain with a $20$~l/s ion pump. We use a $1064$~nm Nd:YAG laser to perform the LITA and release the strontium vapour. The fibre-coupled output of the laser has a single transverse mode, and a maximum power of $38$~W. We focus the ablation laser beam down to a small spot of waist $w_0\simeq4$~$\mu$m to achieve a strong local heating effect. This ensures that the strontium vapour can be released within tens of milliseconds after the ablation laser is turned on. At the same time, the ablation spot is much smaller than the centimetre-sized strontium granule. The granule then acts as a heat reservoir to guarantee that once the ablation laser is turned off, the ablation spot is rapidly cooled down to room temperature, allowing the background pressure to return to its strontium-free value. 


To produce a cloud of cold atoms in a MOT from the released vapour, we use the $461$~nm $^1S_0\rightarrow\,^1P_1$ cooling transition which has a linewidth of $\Gamma/2\pi=32$~MHz. As shown in Figure~\ref{fig:exp setup}, the three counterpropagating pairs of MOT beams are arranged to be mutually orthogonal and intersect at a region that is typically 60~mm away from the granules. The frequency of the MOT beams is set at a detuning of $\delta =-1.2\Gamma$ relative to the resonance frequency of the cooling transition. Each MOT beam has a power of $30$~mW and a beam diameter of $18$~mm. The total peak intensity of all six beams is $35\,\textrm{mW/cm}^2$, which is slightly smaller than the saturation intensity of $43\,\textrm{mW/cm}^2$. This cooling transition is not completely closed, which leads to optical pumping of the atoms to the metastable $^3P_2$ state. To recycle these atoms back to the ground state, we send $679$~nm and $707$~nm repumping lasers along one of the horizontal MOT beam path. The repumping laser are continuously turned on. We check that the MOT fluorescence signal is independent of the repumping laser powers, indicating that we suppress the atom losses due to optical pumping.

\subsection*{Quadruple magnetic field}
\begin{figure}[H]
\centering
\includegraphics[width=0.5\linewidth]{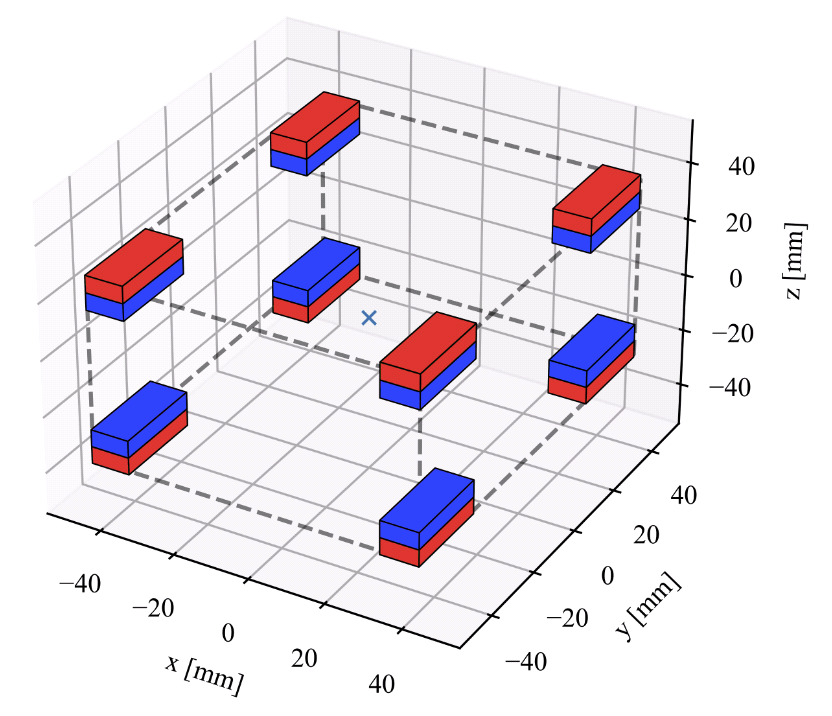}
\caption{Configuration of magnets arranged on eight vertices of a cuboid (four magnets at each vertex). The red and blue colours represent the two poles of the magnets. The magnetic field direction points from blue to red.}
\label{fig:magnetdrawing1}
\end{figure}

To simplify the overall design, we implement a method alternative to the conventional electromagnets to generate the MOT magnetic field gradient. We use permanent magnets as such solutions have already been deployed in 2D MOTs and Zeeman slowers~\cite{lamporesi2013compact,cheiney2011zeeman}. The arrangement of the magnets is shown in Figure \ref{fig:magnetdrawing1}. Thirty-two pieces of neodymium-iron-boron (NdFeB) permanent magnets (N750-RB from Eclipse Magnetic Ltd.) are placed in a cuboid configuration, where each of the eight vertex hosts four magnets. The magnetization is aligned along the short dimension of the magnets. The arrangement gives magnetic field gradients of $\sim 50$~G/cm along the vertical direction and $\sim 25$~G/cm in the horizontal plane. Such values are typically used to operate a $461$~nm strontium MOT~\cite{yang2015high}. In addition, our arrangement of the magnets ensures that a large optical access to the atomic cloud is available.

\subsection*{Data analysis}
We used a CMOS camera to record a temporal series of images of the cold atomic cloud with a frame rate of $50$ frames per second. For each image, we remove the background light contribution that is recorded by an image taken before the ablation laser is turned on. We then fit the recorded fluorescence images by a Gaussian profile with a non-zero offset, which is proportional to the amount of strontium vapour in the glass cell. Thus, the fit allows us to extract the temporal evolution of both the cold atoms number (see the supplementary information for the calculation of number of atoms from MOT fluorescence signal) as well as the amount of strontium vapour in the glass cell. From these temporal profiles, we obtain the loading times and the lifetimes of the cold atoms and of the vapour.

\section*{Results and discussions}

In this section, we report on the performances of our ablation-MOT apparatus. The timing of the ablation laser is controlled from a PC through serial communication. The shortest ablation pulse duration that we can program is $25$~ms. The pulse has a rising and falling time of $\sim1$~ms each. With this technical consideration in mind, we explore a short and a long ablation pulse sequence. The short pulse sequence aims to minimize the granule heating for a fast relaxation of the strontium vapour (few tens of milliseconds), and an optimum lifetime of the cold gas. For that purpose, we use the maximum available power of the ablation laser with a minimal pulse duration. For the long pulse sequence, we aim to trap a larger population of atoms in the MOT. Here, we use a weaker laser power to avoid overheating the granule, leading to a longer loading time of the MOT.



\subsection*{Short pulse sequence}
\label{SP}

\begin{figure}[H]
\centering
\includegraphics[width=0.8\linewidth]{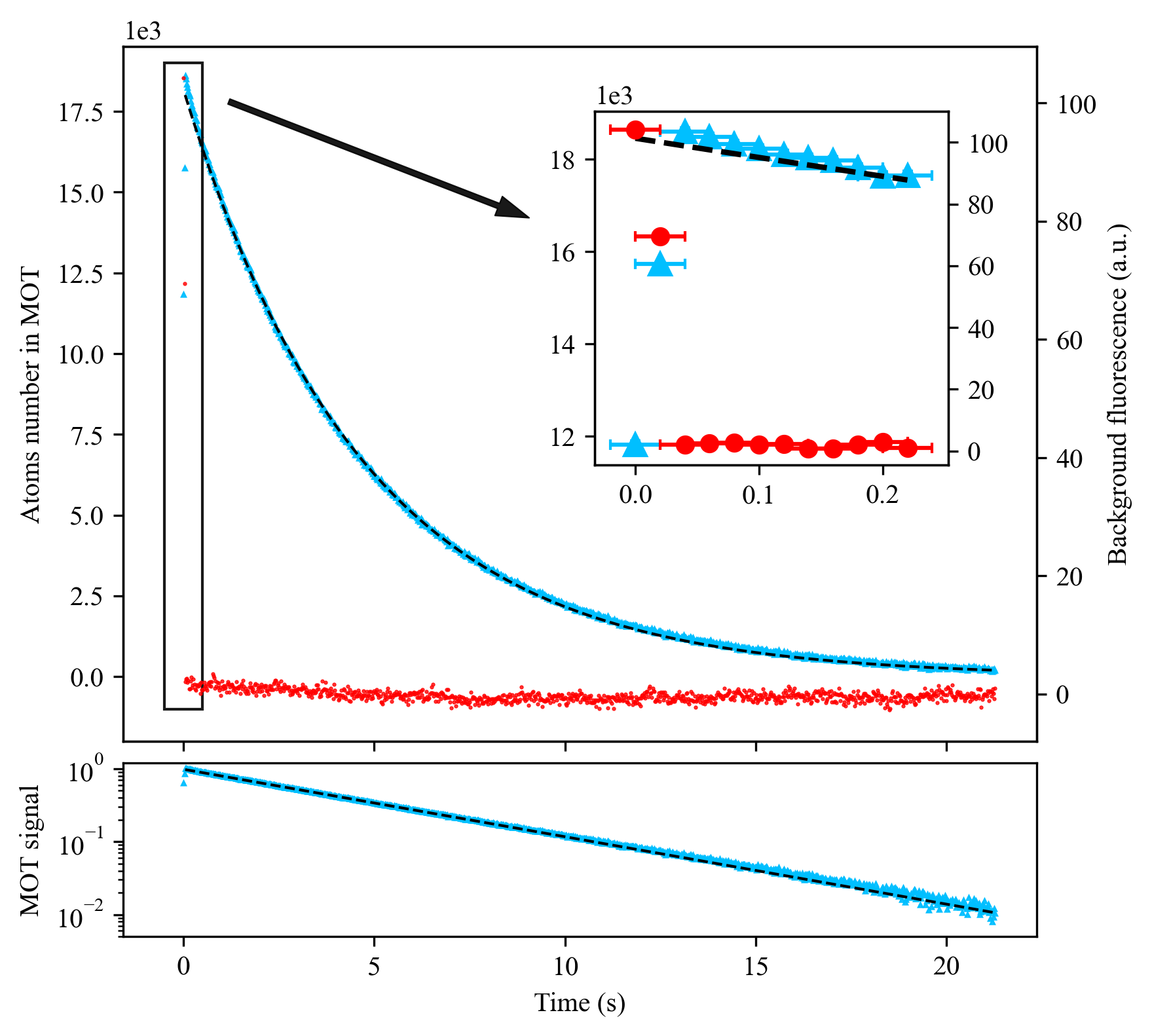}
\caption{Experimental results of the short pulse sequence. \textit{Top panel}: Temporal evolution of the cold atoms number (light blue triangles) and the strontium vapour (red circle), averaged over $78$ runs of the experiment. The time origin corresponds to the ignition of the ablation laser. The decaying part of the MOT signal is fitted to an exponential function (black dashed line) to give a MOT lifetime of $\tau_N= 4.7(9)$~s. The uncertainty reflects one standard deviation of the lifetime measured over 1.5 hours for the 78 runs of experiments. \textit{Inset}: A zoom of the first $200$~ms. The horizontal error bars reflect the temporal resolution of our measurement methods. \textit{Bottom panel}: Normalized temporal evolution of atomic fluorescence signal in semi-log scale. }
\label{fig:shortpulse}
\end{figure}

In the short pulse sequence, the ablation laser is turned on for a duration of $25$~ms, which is the minimum duration allowed by the laser module. We apply a maximal ablation power of $38$~W. 
In Figure~\ref{fig:shortpulse} we plot the temporal evolution of the cold atoms number  (light blue triangles) and the fluorescence signal of the strontium vapour (red circles). We find a loading time of $40(10)$~ms, where the precision is limited by the temporal resolution of our measurement device, and measure a peak value of $1.9(4)\times$10$^4$ atoms in the MOT. Even though the atoms number achieved in this sequence is small as compared to conventional setups~\cite{yang2015high}, it is likely to be sufficient for applications in optical tweezers such as quantum computing involving Rydberg atoms~\cite{saffman2016quantum,cohen2021quantum} and clocks~\cite{Young2020}.

The red dots, showing the background fluorescence, indicate a rapid drop of the strontium vapour to zero with a characteristic time of 40(10)~ms (see inset, Figure \ref{fig:shortpulse}). We understand this fast decrease in strontium vapour for two reasons: First, the heating volume is very small with respect to the size of the granule. In absence of the ablation laser, the granule behaves as a thermal reservoir to reduce rapidly the temperature by thermal conductance. Second, atoms in the vapour are rapidly adsorbed by the chamber walls, contributing to a pressure reduction when we turn off the ablation laser. 
Since the background pressure rapidly relaxes to a strontium-free steady-state value, the cold atoms number should undergo an exponential decay. We confirm this exponential behaviour by plotting the normalized value of the number of atoms in semi-logarithmic scale (see Figure \ref{fig:shortpulse} bottom panel). We find a MOT lifetime of $\tau_N\sim~4.7(9)$~s, consistent with a low background pressure environment.

\subsection*{Long pulse sequence}
\label{LP}

\begin{figure}[H]
\centering
\includegraphics[width=0.8\linewidth]{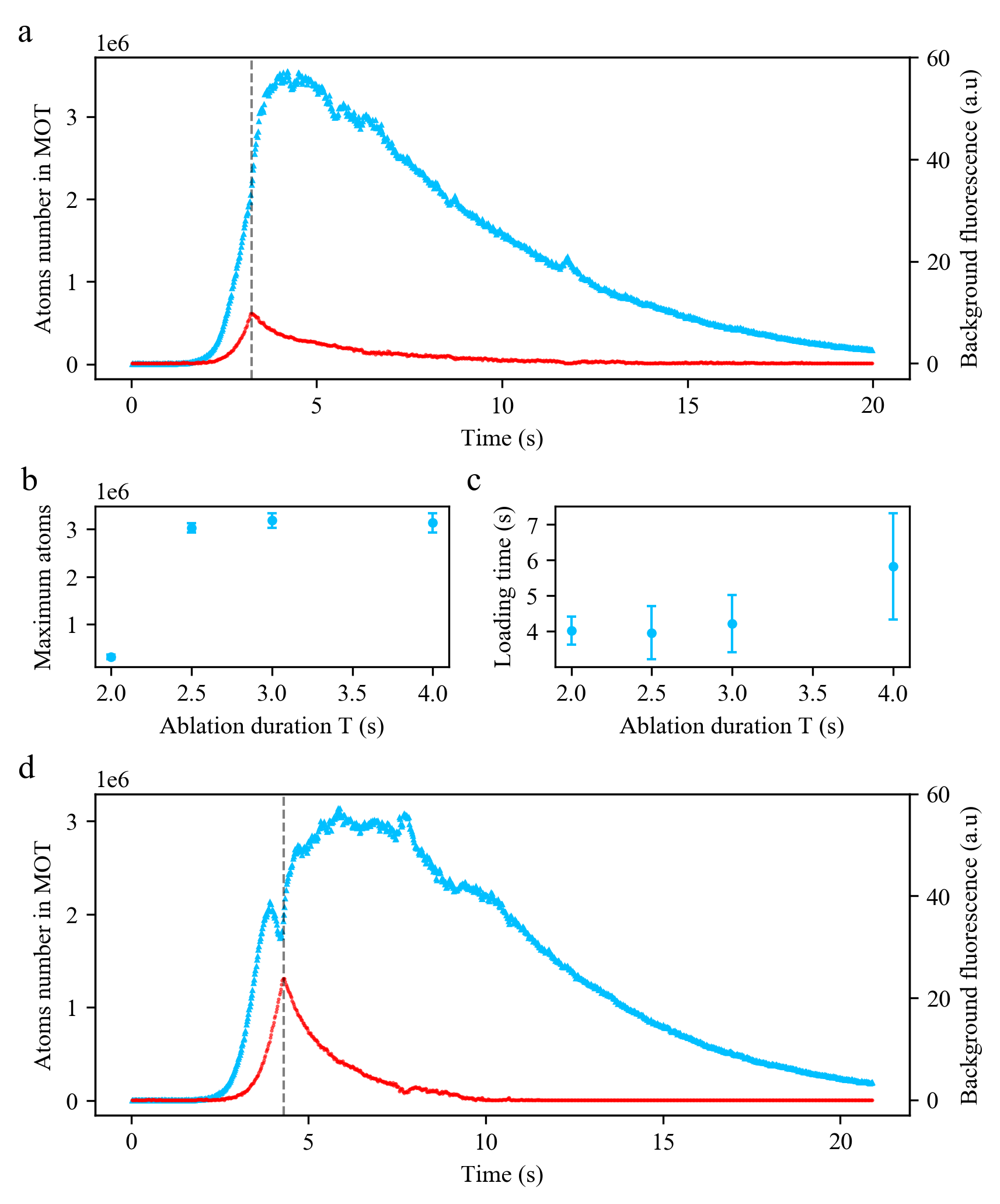}
\caption{Experimental results of the long pulse sequence. \textbf{a.} Temporal evolution of the fluorescence signal of the cold atoms cloud (light blue triangles) and the strontium vapour (red circles) for an ablation duration of $T=3$~s, and a laser power of $20$~W. The vertical dashed line shows the time at which the ablation laser turns off. \textbf{b.} The peak cold atoms number as a function of ablation duration $T$. Fluorescence signals are observed when $T\geq 2$~s. \textbf{c.} Loading times (time taken to reach the peak atoms number value after the ablation laser ignition) as a function of the ablation duration $T$. \textbf{d.} Temporal evolution of the fluorescence signal of the cold atoms cloud (light blue triangles) and the strontium vapour (red circles) for an ablation duration of $T=4$~s, and a laser power of $20$~W. The vertical dashed line indicates the time at which the ablation laser turns off. In a. and d., the time origin corresponds to the laser ablation ignition.}
\label{fig:longpulse}
\end{figure}

For cases where a higher number of atoms is required, we can increase the ablation pulse duration. This point is illustrated in Figure \ref{fig:longpulse}a, where we show the temporal evolution of the cold atoms number  (light blue triangles) and the fluorescence of the background strontium vapour (red circles), for an ablation duration of $T=3$~s, and an ablation power of $20$~W. A lower power is used here, as compared to the short pulse sequence, to avoid overheating of the strontium granule. Under these conditions, we find a peak of $3.5(2)\times$10$^6$ atoms in the MOT at around $4$~s after ignition of the ablation laser. 

The background fluorescence level of the strontium vapour increases following an exponential-like behaviour with a characteristic time of $400(2)$~ms. The peak value of the vapour signal coincides with the time at which the laser turns off (dashed vertical lines). After the laser is turned off, the vapour decreases following again an exponential-like behaviour with a characteristic time of $2.50(1)$~s. The number of atoms in the MOT follows a similar qualitative trend. More precisely, during the ablation, the number of atoms increases with time until it reaches a peak value around one second after the ablation laser is turned off. Due to the long ablation duration, the granule stays hot and the strontium vapour is not removed immediately after the ablation laser is turned off, allowing the loading of the MOT to carry on until the strontium vapour becomes sufficiently low ($\sim 10$~s). At longer times ($10-20$~s), we find an expected exponential decay with a lifetime of $\tau_N= 4.2(9)$~s, in agreement with the lifetime measured in the short pulse sequence.

We systematically study the performance of the long pulse sequence (with laser power at $20$~W) in terms of the peak atoms number and loading time. The loading time is defined as the time taken to reach the peak atoms number value after ignition of the ablation laser. In Figures \ref{fig:longpulse}b and \ref{fig:longpulse}c, we show these values for ablation duration ranging from $T=2$~s to $T=4$~s.  We observe that the peak atoms value saturates for $T\geq 2.5$~s, whereas the loading time increases with $T$. These results suggest that a larger strontium vapour background, obtained with larger $T$, does not help to increase the atoms numbers and to some extent can be detrimental since it increases the loading time.

To get a better understanding of the mechanisms at play at longer pulse duration, we show in Figure \ref{fig:longpulse}d, the temporal evolution of the cold atoms number  (light blue triangles) and the fluorescence of the strontium vapour (red circles) for an ablation duration of $T=4$~s. It shows a striking feature of a dip in the atoms number temporal profile, when the strontium vapour is going to reach its peak value. This indicates that the background pressure becomes too high such that the loss rate due to thermal atom-cold atom collisions surpasses the cooling rate of the MOT. The increasing of atoms in the MOT resumes after switching off the ablation laser, \textit{i.e.} after a reduction of the background pressure.

\section*{Conclusion and perspectives}

We demonstrated a simple and compact approach to operate a magneto-optical trap using the LITA technique as an atomic source. With a  short pulse sequence, we obtain a short loading time of $40(10)$~ms followed by a rapid decay of the strontium vapour to maintain an ultra-high vacuum environment for a subsequent manipulation of the cold gas. With a long pulse sequence, we report up to $3.5$ million cold strontium atoms. Our results serve as a stepping stone towards even more compact and portable setups, illustrating a proof-of-concept that such techniques are highly viable. 

An important merit of our ablation-MOT system is the possibility to directly capture the vaporised atoms into a MOT with a trapping lifetime of more than 4~s. Since there is no need for Zeeman slower and 2D-MOT, a compact setup with less requirement for maintenance can be developed in future. Importantly, the LITA is based on the thermal process of laser heating, and therefore, the actual wavelength for LITA should not be critical. In many cold atom experiments, a high-power laser is used to create a far-off-resonance trap of the cold atoms. This laser could serve as the ablation laser for LITA, thus removing the need to maintain another laser.

We performed the MOT loading with a constant ablation laser power, but it might not be the best experimental strategy for applications requiring larger cold atoms number. Instead, one can think of a MOT loading sequence at a constant strontium vapour pressure. To realize it, one should implement, for example, a decreasing temporal profile on the ablation laser power.

We envision the potential applications of the ablation-MOT setup for outdoor precision-measurement experiments. The simple source of ultracold strontium also finds potential applications in quantum computing and simulations using optical lattices and tweezers. Lastly, the ablation technique opens up unprecedented laser cooling possibilities to other elements with low saturation pressure at room temperature.



\section*{Acknowledgements}

This work was supported by the CQT/MoE funding Grant No. R-710-002-016-271, and by the Singapore Ministry of Education Academic Research Fund Tier2 Grant No. MOE2018-T2-1-082 (S).

\section*{Author contributions statement}

C.C.H., C.C.K, and D.W. developed the experimental apparatus and analysed the data. C.C.H. took the data. R.L. devised the first ablation system. All authors reviewed the manuscript and contributed to the writing.

\section*{Additional information}

\subsection*{Competing interests}
The author(s) declare no competing interests.



\section*{Supplementary material}

# Supplementary Information: Laser-induced thermal source for cold atoms

Chung Chuan Hsu[1], Rémy Larue[2,3], Chang Chi Kwong[1,3,*], and David Wilkowski[1,3,4]

[1]School of Physical and Mathematical Science, Nanyang Technological University, 637371 Singapore, Singapore
[2]Laboratoire de Physique Subatomique et de Cosmologie, Université Grenoble-Alpes, CNRS/IN2P3, Grenoble INP, 38000 Grenoble, France
[3]MajuLab, International Research Laboratory, IRL 3654, CNRS, Université Côte d'Azur, Sorbonne Université, National University of Singapore, Nanyang Technological University, Singapore.
[4]Centre for Quantum Technologies, National University of Singapore, 117543 Singapore, Singapore
*changchikwong@ntu.edu.sg

Here, we describe how the number of atoms in the MOT is approximated from the fluorescence signal captured by a system with a detection solid angle of $\Omega$. Our work is done only on the $^{88}$Sr isotope. The cooling transition is a $J = 0 \to J = 1$ transition, where scattering cross section is the same for all laser polarizations. We start by considering a single atom and calculate $P_a$, which is the fluorescence power emitted by a single strontium atom into the solid angle of $\Omega$:

$$P_a = E_p \Gamma' \frac{\Omega}{4\pi} \tag{1}$$

Here, $E_p = hc/\lambda$ is the energy of a photon at $\lambda = 461$ nm, and $\Gamma'$ denotes the rate of number of photons scattered per atom. We denote the speed of light by $c$ and Planck's constant by $h$. We consider $P_a$ to be averaged over many fluorescence events of all the atoms in the MOT and assume that its angular distribution is isotropic. Since the natural linewidth is $\Gamma = 32 \times 2\pi$ MHz,

$$\Gamma' = \frac{1}{2}\frac{s}{1+s}\Gamma = \frac{1}{2}\frac{s_0}{4\delta^2/\Gamma^2 + 1 + s_0}\Gamma \tag{2}$$

where $s = s_0/(4\delta^2/\Gamma^2 + 1)$, $\delta$ is the detuning of the MOT beams, $s_0 = I_L/I_S$ is the saturation parameter and $I_S$ is the saturation intensity of the cooling transition, which is 43 mW/cm$^2$. For our MOT beams with a total power of $P$ and a waist of $w$, the total intensity is $I = 2P/\pi w^2$. For small detection angles, we perform the approximation $\Omega/4\pi \approx \frac{1}{2}(1 - \cos\frac{D}{2d})$, where the camera lens diameter is $D$ and the imaging distance is $d$. This leads to the following expression for $P_a$

$$P_a = E_p\Gamma'\frac{\Omega}{4\pi} = \left(\frac{hc}{\lambda}\right)\left(\frac{1}{2}\frac{s_0}{4\delta^2/\Gamma^2 + 1 + s_0}\Gamma\right)\left[\frac{1}{2}\left(1 - \cos\frac{D}{2d}\right)\right] \tag{3}$$

Finally, we divide the total MOT power detected by the camera $P_T$ by $P_a$ such that number of atoms is therefore $N = P_T/P_a$. $P_T$ was pre-determined via calibrating the counts to a known beam power. Using this formula, we are able to estimate the number of atoms for any given fluorescence data captured by the CMOS camera in our experiments.



\end{document}